\documentclass[fleqn,usenatbib]{mnras}

\usepackage{newtxtext,newtxmath}

\usepackage[T1]{fontenc}
\usepackage{ae,aecompl}

\usepackage{graphicx}	
\usepackage{amsmath}	
\usepackage{amssymb}	
\usepackage{gensymb}
\usepackage{rotating}
\usepackage{float}
\restylefloat{table}
\restylefloat{figure}
\usepackage{enumerate}
\usepackage[utf8]{inputenc}
\usepackage{caption} 
\usepackage{verbatim}
\usepackage{lineno}

\title[A search for $\gamma$-ray emission from nearby radio galaxies.]{A search for $\gamma$-ray emission from a sample of local universe low-frequency selected radio galaxies.}

\author[M. Harvey et al.]{
Max Harvey,$^{1}$\thanks{E-mail: max.harvey@durham.ac.uk}
Cameron B. Rulten$^{1}$ and
Paula M. Chadwick$^{1}$
\\
$^{1}$Centre for Advanced Instrumentation, Department of Physics, University of Durham, South Road, Durham DH1 3LE, United Kingdom
}

\date{Accepted XXX. Received YYY; in original form ZZZ}

\pubyear{2019}

\begin{document}
\label{firstpage}
\pagerange{\pageref{firstpage}--\pageref{lastpage}}
\maketitle

\begin{abstract}
Radio galaxies are uncommon $\gamma$-ray emitters, and only low redshift radio galaxies are detected with \textit{Fermi}-LAT. However, they offer potential insights into the emission mechanisms of active galaxies, particularly as the alignment of their jets with respect to the Earth means that, unlike blazars, their emission is not necessarily jet-dominated. We use \textit{Fermi}-LAT data to perform an unbiased survey of 78 radio galaxies from the Bologna Complete Sample in order to search for new $\gamma$-ray-emitting radio galaxies. We observe statistically significant $\gamma$-ray emission from 4 of the 6 known \textit{Fermi}-LAT detected radio galaxies included in this sample, and find some evidence for $\gamma$-ray emission spatially coincident with 4 previously undetected radio galaxies. As a large parameter space is searched, we calculate a probability distribution to compute the look-elsewhere effect. We find that these 4 spatially coincident sub-threshold $\gamma$-ray excesses are most likely a chance association, and are unlikely to be emission from the radio galaxies. Upper limits on flux are calculated for the radio galaxies from which no $\gamma$-ray emission is observed. 
\end{abstract}





\section{Introduction} 

\subsection{Radio Galaxies}

Active galactic nuclei (AGN) can be divided into two broad classes: radio-quiet AGN, which have little to no radio emission relative to other wavebands, and radio-loud AGN, which have dominant radio emission relative to other wavebands. It is the radio-loud AGN, primarily blazars and radio galaxies, that are the most prominent in the $\gamma$-ray sky. Both have powerful radio jets, up to kiloparsec (kpc) scales in many cases, and are considered to be fundamentally the same objects viewed at different orientation angles. 

Blazars are radio-loud AGN which are viewed nearly directly down the jet, from which we see variable and luminous radio emission dominated by Doppler boosting of the electromagnetic radiation in the jet. In the $\gamma$-ray sky, over 90\% of extragalactic objects detected are blazars, i.e. BL~Lacs and Flat Spectrum Radio Quasars (FSRQs) (\citealt{massaro_extragalactic_2016}, \citealt{the_fermi-lat_collaboration_fermi_2019}). There is a general consensus that the $\gamma$-rays are produced through inverse Compton scattering by electrons in the jet, but other mechanisms are possible (e.g. \citealt{boettcher_leptonic_2013}). However, most AGN are not viewed from such a small angle; the bulk of the radio loud population, the radio galaxies (\citealt{urry_unified_1995}, \citealt{ atwood_large_2009}), is viewed from a larger orientation angle relative to the jet, so that the observed emission is not necessarily dominated by Doppler boosting in the jet.

The larger orientation angles of radio galaxies allow us to view their extended structure, thus we are able to define some differences between two types of kpc scale jet: Fanaroff-Riley I (FR\,I) radio galaxies have `low power' sub-relativistic kpc scale jets, which are radiatively inefficient, edge darkened and terminate in radio plumes. On the other hand, Fanaroff-Riley~II (FR\,II) radio galaxies have `high power' mildly relativistic kpc scale jets, are more radiatively efficient than FR\,I galaxies, and are edge brightened, terminating in radio lobes containing hot-spots of radio emission \citep{fanaroff_morphology_1974}. On a parsec scale, close to the radio core, the jets of FR\,I and FR\,II radio galaxies are identical \citep{giovannini_vlbi_1994}. It is thought that FR\,I radio galaxies are the parent population of the BL Lac blazar class, and FR\,IIs are the parent population of FSRQs (\citealt{grandi_exploring_2012}, \citealt{urry_unified_1995}).

\subsection{Gamma-Ray Observations}

Launched in 2008, the \textit{Fermi} Large Area Telescope (\textit{Fermi}-LAT), has been observing the whole sky over an effective $\gamma$-ray energy range of 100\,MeV to approximately 300\,GeV for over a decade. Varying levels of $\gamma$-ray emission have been detected in both FR\,I and FR\,II radio galaxies (\citealt{abdo_fermi_2009}, \citealt{rulten_search_2020}). There is evidence for $\gamma$-ray production in both the radio core of the galaxy, from the parsec scale jet \citep{angioni_gamma-ray_2019}, and the kpc scale jets for both galaxy types (\citealt{abdo_fermi_2010}, \citealt{grandi_exploring_2012}) 

There is also evidence for $\gamma$-ray emission from a third type of radio galaxy, the FR\,0, or `compact' radio galaxy (\citealt{baldi_radio_2009}, \citealt{baldi_new_2016}). A relatively recent addition to the FR classification system, the FR\,0 radio galaxy is defined by a core to total emission ratio 30 times that of the FR\,I, and is believed to be an early evolutionary phase in the activity of an FR\,I radio galaxy which is lacking extended radio emission in kpc jets \citep{garofalo_fr0_2019}. Despite the FR\,0 class being the most numerous type of radio galaxy in the local universe, almost all of the radio galaxies with detected $\gamma$-ray emission are FR\,I or FR\,II in nature.

\subsection{Spectral Features in AGN and The Bologna Complete Sample}

The Bologna Complete Sample (BCS) (\citealt{giovannini_bologna_2005}, \citealt{liuzzo_bologna_2009}) is a catalogue of 95 radio-loud AGN. The BCS contains two BL~Lac blazars, Mrk\,421 and Mrk\,501, and 93 radio galaxies of varying kpc scale jet morphology. Of the 95 galaxies in the complete catalogue, 8 are included in the 4th \textit{Fermi} Point Source Catalogue (4FGL) \citep{the_fermi-lat_collaboration_fermi_2019}: Mrk\,421 and Mrk\,501, the FR\,I galaxies NGC\,315, NGC\,2484, 3C\,264, and 3C\,274 (M\,87 / Virgo\,A), 4C\,29.41 and the FR\,0 galaxy 4C\,39.12.

The BCS is a particularly interesting sample for investigating the characteristics of $\gamma$-ray emission from radio galaxies. It is drawn from the 3CR \citep{bennett_preparation_1962} catalogue (178 MHz) and B2 catalogue (408 MHz) (\citealt{colla_catalogue_1970}, \citealt{colla_b2_1972}, \citealt{colla_b2_1973} and \citealt{fanti_b2_1974}). The authors impose no selection criteria on core radio power, orientation angle, or the velocity of the jet. The primary selection criteria are a redshift limit ($z < 0.1$), and a cut on Galactic latitude, keeping only sources where $|b| > 10\degree$. The precursor to the BCS, the Bologna Strong Core Sample (BSCS) (\citealt{giovannini_vlbi_1990} and following works), had a core flux limit which meant that the BSCS was inherently biased towards more Doppler boosted sources with smaller orientation angles and therefore dominated by emission from the kpc scale jet. The BCS, however, has no such limit, and consequently has a range of sources with a variety of orientations. 

Most detected $\gamma$-ray emitting radio galaxies are low redshift objects (z < 0.1), therefore we expect that any new radio galaxy detections with \textit{Fermi}-LAT would have similarly low redshifts. In order to check that the \textit{Fermi}-LAT detected radio galaxies and the BCS have similar redshift distributions,  we employ a 2-sample Kolmogorov-Smirnov (KS) test\footnote{The equation for the 2-sample KS test is given as $D_{a, \: b} = \sup |F_{1, \: a(x)} - F_{2, \: b(x)}|$ for two samples $a(x)$ and $b(x)$, where $D_{a, \: b}$ is the KS statistic.} between the 4FGL $\gamma$-ray emitting radio galaxies and the BCS radio galaxies. We find a p-value of $p = 0.131$, indicating reasonable similarity between the 4FGL radio galaxies' redshifts and those of the BCS objects (Table \ref{tbl:radgals} in the Appendix).

There is also a connection between the flux levels of the radio and optical wavebands in radio galaxies \citep{owen_fri/il_1994}. A 2-sample KS test comparing the observed absolute magnitudes of the BCS radio galaxies with those of the \textit{Fermi}-LAT detected radio galaxies gives a p-value of $p=0.949$, indicating that the optical luminosity distribution of the radio galaxies in the BCS is similar to that of the $\gamma$-ray radio galaxies.

The closest AGN to Earth, the FR\,I radio galaxy Centaurus A (Cen\,A), shows an unusual spectral feature in $\gamma$-rays. Due to its proximity, Cen\,A is one of two radio galaxies where significantly extended $\gamma$-ray emission is seen with \textit{Fermi}-LAT \citep{abdo_fermi_2010}, the other being Fornax\,A \citep{ackermann_fermi_2016}. In Cen\,A, the Fermi-LAT data show a distinct $\gamma$-ray core and lobe structure consistent with the positions of the radio core and lobes. \cite{brown_discovery_2017} analyzed the core emission, and found a statistically significant ($5 \sigma$) hardening in the \textit{Fermi}-LAT spectrum, with a break energy of $2.6 \pm 0.3$\,GeV. This spectral feature is compatible with a `spike' in the dark matter halo profile, but could also be produced by a population of unresolved millisecond pulsars or signatures of cosmic ray acceleration. All of these hypotheses are consistent with $\gamma$-ray emission that does not originate in the kpc scale jet. Recent observations with the H.E.S.S. $\gamma$-ray telescope facility suggest that the highest energy emission may in fact come from an inner parsec scale jet, close to Cen\,A's core \citep{sanchez_morphology_2018}. This result is consistent with the results of \cite{angioni_gamma-ray_2019}, where they find that pc scale jets drive the $\gamma$-ray emission in galaxies which are not dominated by Doppler boosted emission.

\cite{rulten_search_2020} examined 26 \textit{Fermi}-LAT detected radio galaxies for Cen\,A-like spectral features, but found no evidence of spectral hardening in the 10 year average spectra among the radio galaxies which showed no evidence for variability. In this work we search for $\gamma$-ray emission from a sample of close, low-frequency radio galaxies which are unlikely to be dominated by Doppler-boosted $\gamma$-ray emission (blazar-like emission). This would constitute evidence for large-scale emission processes which are not associated with a jet occurring in these objects.

In Section\,\ref{sec:select} we discuss the selection of the BCS galaxies discussed in this paper, and in Section\,\ref{sec:analysis} we describe our analysis of the \textit{Fermi}-LAT data. Our results are given in Section\,\ref{sec:results}, and our statistical analysis is provided in Section\,\ref{sec:stats}. Finally, some discussion and conclusions are given in Sections\,\ref{sec:discussion} and \ref{sec:conclusions} respectively.

\section{Selection Criteria} 
\label{sec:select}

Not all of the 95 galaxies in the 2005 edition of the BCS \citep{giovannini_bologna_2005} are suitable for our search with \textit{Fermi}-LAT. The two Markarian objects are immediately discarded as they are blazars, and thus not relevant to our work. We discard any radio galaxy in the BCS without a clear `FR\,I', `FR\,II' or `C' (FR\,0) morphology since it would be difficult to draw any conclusions about such objects. The BCS authors' cuts on Galactic latitude and redshift are sufficient for our purposes, so we do not impose any further cuts. We also do not impose any cuts on radio luminosity, core radio power, or any other property at any wavelength. 


We use the 6 objects with known $\gamma$-ray emission, which are included in the 4FGL, as a reference for the analysis of any newly-detected sources rather than considering them within our search. Finally, several sources were discarded after initial analysis of the \textit{Fermi}-LAT data, as the modelling close to the position of the target radio galaxy was of poor quality (for instance due to very bright sources in the field of view). After applying our selection criteria, we are left with 72 target radio galaxies from which we attempt to find evidence of $\gamma$-ray emission, plus the 6 known 4FGL radio galaxies for reference. A full list of the galaxies included in the sample (including those with known $\gamma$-ray emission), with their morphology, coordinates and redshift, is shown in Appendix Table \ref{tbl:radgals}. 

For the purposes of this paper, we will refer to the target radio galaxies in our sample as TRGs, and those where a level of $\gamma$-ray emission is observed that is below the $5 \sigma$ threshold for detection, as sub-threshold $\gamma$-ray target radio galaxies (STRGs). 

\section{\textit{Fermi}-LAT Observations and Analysis} 
\label{sec:analysis}

The \textit{Fermi}-LAT \citep{atwood_large_2009} is a pair conversion telescope which detects the charged particles produced by $\gamma$-rays interacting with the detector, from which it is possible to reconstruct the energy, time of arrival and incident direction of the photon. We take all-sky observations over an 11.5 year period to execute our survey, with the parameters used in our observations and modelling shown in Table \ref{tbl:parameters}. The 11.5 period is chosen to maximise observation time, and runs from the beginning of the \textit{Fermi}-LAT mission until the week when we began our analysis in January 2020.

\begin{table}[!ht]
\centering
\begin{tabular}{cc}
\hline \hline
Observation Period (Dates) & 04/08/2008 - 10/01/2020 \\
Observation Period (MET) & 239557417 - 600307205 \\
Observation Period (MJD) & 54682 - 58858 \\
Energy Range (GeV) & 0.1 - 300 \\
Data ROI width & $20\degree$ \\
Model ROI Width & $30\degree$ \\
Zenith Cut & $< 90\degree$ \\
Instrument Response  & \texttt{P8R3\_SOURCE\_V2} \\
Point Source Catalogue & 4FGL \\
Isotropic Diffuse & \texttt{iso\_P8R3\_SOURCE\_V2\_v01} \\
Galactic Diffuse & \texttt{gll\_iem\_v07} \\
\hline
\end{tabular}
\caption{The parameters used in the likelihood analysis of the square ROI around each TRG.}
\label{tbl:parameters}
\end{table}

We use Pass 8 \textit{Fermi}-LAT data, which has improved analysis methods and event reconstruction over prior data-sets. Using the parameters in Table \ref{tbl:parameters}, we execute a standard reduction chain using the \textit{Fermi} Science Tools: energy cuts on photons, computing instrument live-time etc. For each TRG, we consider a square region of interest (ROI) of $20 \degree$ width around it, and bin all photons into spatial bins of $0.1 \degree$ width in RA and Dec, and in 10 bins per decade for energy. We follow the well-established binned likelihood analysis method of \cite{mattox_likelihood_1996}, whereby maximum likelihood estimation is used to fit a model to our data-set on a bin-by-bin basis. We use the 4FGL catalogue and background model components described in Table \ref{tbl:parameters}, to make predictions for the number of photons in each bin. To improve the accuracy of the predictions in the model, we first iteratively push the parameters close to the maximum likely values for each catalogued source in the model. We then free the spectral normalisation of all sources in each ROI within $3 \degree$ of the central, target radio galaxy, and free the normalisation of both the isotropic and Galactic background components. We then perform a full maximum likelihood fit on the sources with freed normalisation. To check our modelling, we compute a residual map, the difference between the observed and predicted counts on a bin by bin basis. Our residual map of the ROIs is shown in Appendix Figure \ref{fig:res_map}.


The majority of known $\gamma$-ray emitting radio galaxies are point sources when viewed with \textit{Fermi}-LAT. The test statistic (TS), used to measure source significance, is defined as the ratio between the likelihood of an alternative ($\Theta_1$), and a null hypothesis ($\Theta_2$), given by Equation \ref{eqn:TS}:
\begin{equation}
    \label{eqn:TS}
    \mathrm{TS} = 2 \ln \frac{L(\Theta_{1})}{L(\Theta_{2})}
\end{equation}
The null hypothesis is that there is no point source, and the alternative hypothesis is that there is one. Via Wilks' Theorem \citep{wilks_large-sample_1938}, the TS corresponds to a value of $\chi ^{2}$ for $k$ degrees of freedom. 

We execute our survey by running the `find sources' algorithm, included in the \textit{Fermipy} Python module \citep{wood_fermipy:_2017}. This fits a test source at each point in our model, and calculates the TS of each of these. Where the TS corresponds to a significance of $3 \sigma$ or greater, we add this permanently to our model. For a $\mathrm{TS} < 25$, we have 3 degrees of freedom (RA, dec and spectral normalisation) because the algorithm iterates over RA and dec in order to find new sources not included in the 4FGL. For $\mathrm{TS} > 25$, we have 4 degrees of freedom, as we fit spectral shape too. 

For a detection, we require a source significance greater than $5 \sigma$. Sources added to the model with significance $ 3 < \sigma < 5$, are not considered to be significantly detected $\gamma$-ray sources. Instead we consider these to be `sub-threshold' emission radio galaxies (STRGs), where there is a possibility of $\gamma$-ray emission, but the statistical significance is marginal.


 Radio galaxies are generally variable on timescales of months \citep{kataoka_-ray_2010}. NGC 1275 is one exception, with variability timescales of hours, but is a very luminous $\gamma$-ray source \citep{tanada_origins_2018}. Another example would be the rapidly variable 3C\,120, one of the more blazar-like objects in the $\gamma$-ray emitting radio galaxy population \citep{janiak_application_2016}. The majority of the population is less luminous in $\gamma$-rays, rendering flux variability on daily timescales undetectable with current instrument sensitivity. Nonetheless, it is possible that the radio galaxies in our sample might be detectable over a short timescale (e.g. a month) but not when averaged over our observation time. Thus, we produce a light-curve at the position of each radio galaxy, using monthly binning. We then calculate a TS for variability, and use this to see whether any of our sample is variable.

 Should we detect a source with a significance over $5 \sigma$, the next step would be to compute a Spectral Energy Distribution (SED) and then attempt to model any spectral features.

\section{Results} 
\label{sec:results}
Sub-threshold (below $5 \sigma$) $\gamma$-ray excesses are found spatially coincident with 4 of the 72 radio galaxies in our sample, giving us 4 STRGs: B2\,0800+24, B2\,1040+31, 3C\,293 and 3C\,272.1. Of these, only one has $\mathrm{TS} >25$, the typical TS threshold used for a source to be catalogued in the 4FGL \citep{the_fermi-lat_collaboration_fermi_2019}. This radio galaxy is B2\,0800+24, and it is discussed further in Section \ref{sec:0800}. For comparison, the six BCS radio galaxies which are known $\gamma$-ray emitters (in the 4FGL) were also modelled over an 11.5 year period. Four of these had a significance of above $5 \sigma$, the exceptions being NGC\,2484 and 4C\,29.41. While these two 4FGL radio galaxies have previously been significant enough to be included in the Fermi-LAT point source catalogues, over an 11.5 year observation time their TS has decreased, when compared to a shorter observation period. The results for all 10 galaxies are displayed in Table \ref{tbl:results}.

\begin{table}[!ht]
\centering
\begin{tabular}{lccl}
\hline \hline
Galaxy & Morphology & 4FGL? & TS \\
\hline 
B2\,1040+31 & FR\,0 & No & 20.47 ($3.64\sigma$)  \\
3C\,272.1 & FR\,I & No & 18.97 ($3.45\sigma$)  \\
3C\,293 & FR\,I & No & 20.33  ($3.62 \sigma$) \\
B2\,0800+24 & FR\,I & No & 28.47 ($4.26\sigma$) \\
4C\,39.12 & FR\,0 & Yes & 86.19 ($> 5 \sigma$) \\
4C\,29.41  & FR\,I & Yes & 20.85 ($3.7\sigma$) \\
NGC\,315 & FR\,I & Yes & 106.3 ($> 5 \sigma$) \\
NGC\,2484 & FR\,I & Yes & 26.63 ($4.07 \sigma$) \\
3C\,274 (M 87) & FR\,I & Yes & 1813 ($ \gg 5 \sigma$) \\
3C\,264 & FR\,I & Yes & 165.8 ($> 5 \sigma$) \\
\hline
\end{tabular}
\caption{The 10 radio galaxies in our sample which show evidence for $\gamma$-ray emission. Most of the galaxies previously detected in the 4FGL are robustly detected in this analysis. The significance values quoted here are calculated using the given number of degrees of freedom for each 4FGL radio galaxy or STRG. The Fanaroff-Riley galaxy morphologies are taken from \citealt{giovannini_bologna_2005}, where those described as `compact' radio galaxies in that paper are described as FR\,0 radio galaxies here.
}
\label{tbl:results}
\end{table}

The lightcurves provide no evidence for significant $\gamma$-ray flaring or variability in our TRGs. For each light-curve a TS for temporal variability is calculated which corresponds to a $\chi^{2}$ value. For STRGs, $\chi^{2}_{reduced} > 1$ in all cases for the null hypothesis that there is no $\gamma$-ray variability. This indicates that the best model to fit the data is one with no variability in time. For TRGs with no $\gamma$-ray emission, $\chi^{2}_{reduced}$ < 1 which indicates that there is insufficient data to fit the model, therefore no evidence for variability.

As the 4 STRGs are all low in significance, it is not possible to perform any spectral analysis, and we interpret these results as 95\% upper confidence limits, which are calculated on the basis of an $\mathrm{E^{2}\; \times}$ differential flux. For the case of the 68 non-detections, upper limits are also calculated at the radio galaxies' positions. These upper limits are shown in Appendix Table \ref{tbl:radgals}.

\subsection{The Marginal Detection of B2\,0800+24}
\label{sec:0800}

Of the 4 STRGs, the only one that could be considered a possible detection is B2\,0800+24, with a TS of 28.47. With 4 statistical degrees of freedom between the hypothesis of there being a $\gamma$-ray point source and the null hypothesis that there is not, this gives a $4.26 \sigma$ significance, which meets the criterion for inclusion into the 4FGL \cite{the_fermi-lat_collaboration_fermi_2019}. In our survey, we regard $5 \sigma$ as the threshold for detection, primarily because of the large population of $< 5 \sigma$ sources, shown in Figure \ref{fig:ts_hist}. Our calculations in Section \ref{sec:LE} demonstrate that sources below $5 \sigma$ are likely to be caused by fluctuations in the $\gamma$-ray background.

\section{Statistical Analysis} 
\label{sec:stats}

\subsection{The Look-Elsewhere Effect}
\label{sec:LE}
We find 4 STRGs which may provide evidence for emission. However, we must ask the question, what is the probability that this $\gamma$-ray excess is just a fluctuation in the background that happens to resemble faint $\gamma$-ray emission from a STRG due to its position in the sky? This is the look-elsewhere effect, which is a phenomenon whereby a false positive result may arise as a direct result of searching a very large parameter space \citep{lyons_open_2008}. Given we are searching in 72 places in the sky, we must quantify the look-elsewhere effect in order to work out the probability that our STRG observations are a false positive. 

Our criterion for spatial coincidence was that the 99\% positional uncertainty ($r_{99\%}$) must be larger than the angular offset of the centre of the $\gamma$-ray emission from the TRG. Thus, it is possible to calculate the area of the sky covered by all of our detected point source positional uncertainties ($A_{PS}$). This is given by Equation \ref{eqn:area}:
\begin{equation}
\centering
A_{PS} = \sum \pi r_{99\%}^{2}.
\label{eqn:area}
\end{equation}
Dividing this area by the total area analyzed ($A_{tot}$), and accounting for overlaps between ROIs, gives us the probability of there being a point source of at least $3 \sigma$ significance at any point in the sky (we will call this event $x$), i.e.:
\begin{equation}
P(x) = \frac{A_{PS}}{A_{tot}}.
\label{eqn:prob}
\end{equation}

\begin{figure}[!ht]
    \centering
    \includegraphics[width=250pt]{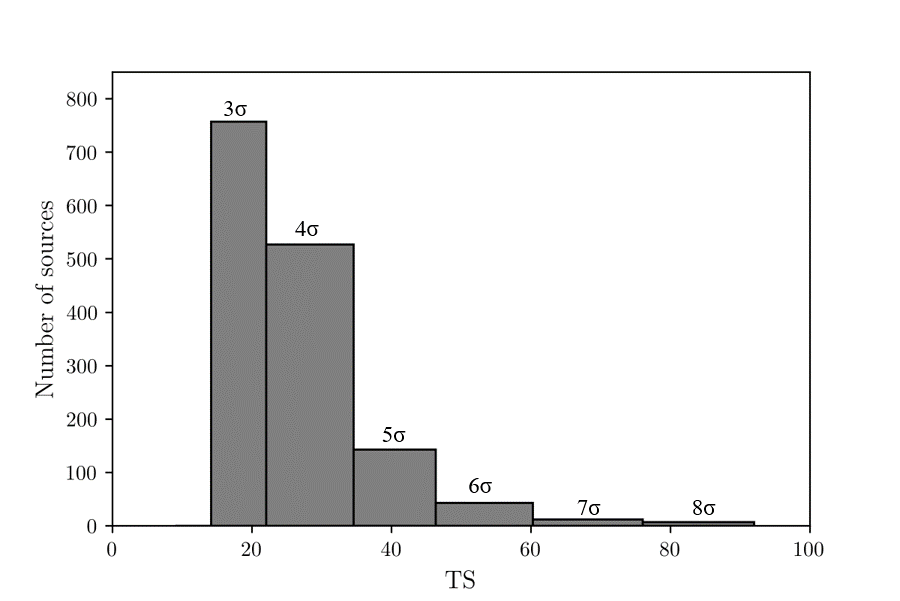}
    \caption{A histogram showing the TS values of all of the non catalogued point sources modelled across all of the 78 ROIs. From left to right the bins show the number of sources of $3 \sigma$, $4 \sigma$, $5 \sigma$, $6 \sigma$, $7 \sigma$ and $8 \sigma$ significance, for the appropriate degrees of freedom in our likelihood analysis. We use the $3 \sigma$ and $4 \sigma$ sources (the two leftmost bins in this distribution) for our look-elsewhere effect calculation, as these represent the significances of the STRGs.}
    \label{fig:ts_hist}
\end{figure}

For our population of point sources across the whole area we have analyzed, we find $P(x) = 0.027$. Using this probability, we can now calculate the probability that 4 STRGs are found, given we looked in 72 places. To do this, we look at the distribution of point sources across the sky, and select all of those with $3 \sigma < \sigma_{PS} < 5 \sigma$, where $\sigma_{PS}$ is the point source significance. The TS distribution of all point sources across all our ROIs is shown in Figure \ref{fig:ts_hist}.

Using the distribution shown in Figure \ref{fig:ts_hist}, we combine the $3 \sigma$ and $4 \sigma$ bins and look at the spatial distribution of these point sources. Employing our false positive probability, $P(x) = 0.027$ (which was calculated for $3 \sigma < \sigma_{PS} < 5 \sigma$ point sources), we set $P(x) = p$ for convenience. We then define our total number of search points in the sky, $N = 72$ and assume that these are distributed randomly across our total analyzed area. We also assume that there is no underlying pattern in the spatial distribution of the point sources; as we are looking solely at the extra-galactic sky, this is a valid assumption to make. Next we define a new variable, $y$, which represents the number of point sources in our TS range which are spatially coincident with our TRGs. For our results, $y = 4$. However, by varying $y$ from $0 \to 72$, we are able to calculate a full probability distribution for the number of chance spatial correlations between TRGs and $3 \sigma < \sigma_{PS} < 5 \sigma$ point sources on the basis of the probability mass function for the binomial distribution\footnote{Although we do not actually replace our sources, with approximately 2000 point sources in our model, the value of $P(x)$ does not significantly change with each association. If we had a number of successful spatial associations that was comparable with our overall sample size, the probability mass function for the hyper-geometric distribution would be more appropriate.}, as we are effectively modelling the success rate of sampling with replacement, given in Equation \ref{eqn:prob_dist}: 

\begin{equation}
P(y) = \binom{N}{y} p^{y} (1 - p)^{(N-y)},
\label{eqn:prob_dist}
\end{equation}

\begin{figure}[!ht]
    \centering
    \includegraphics[width=240pt]{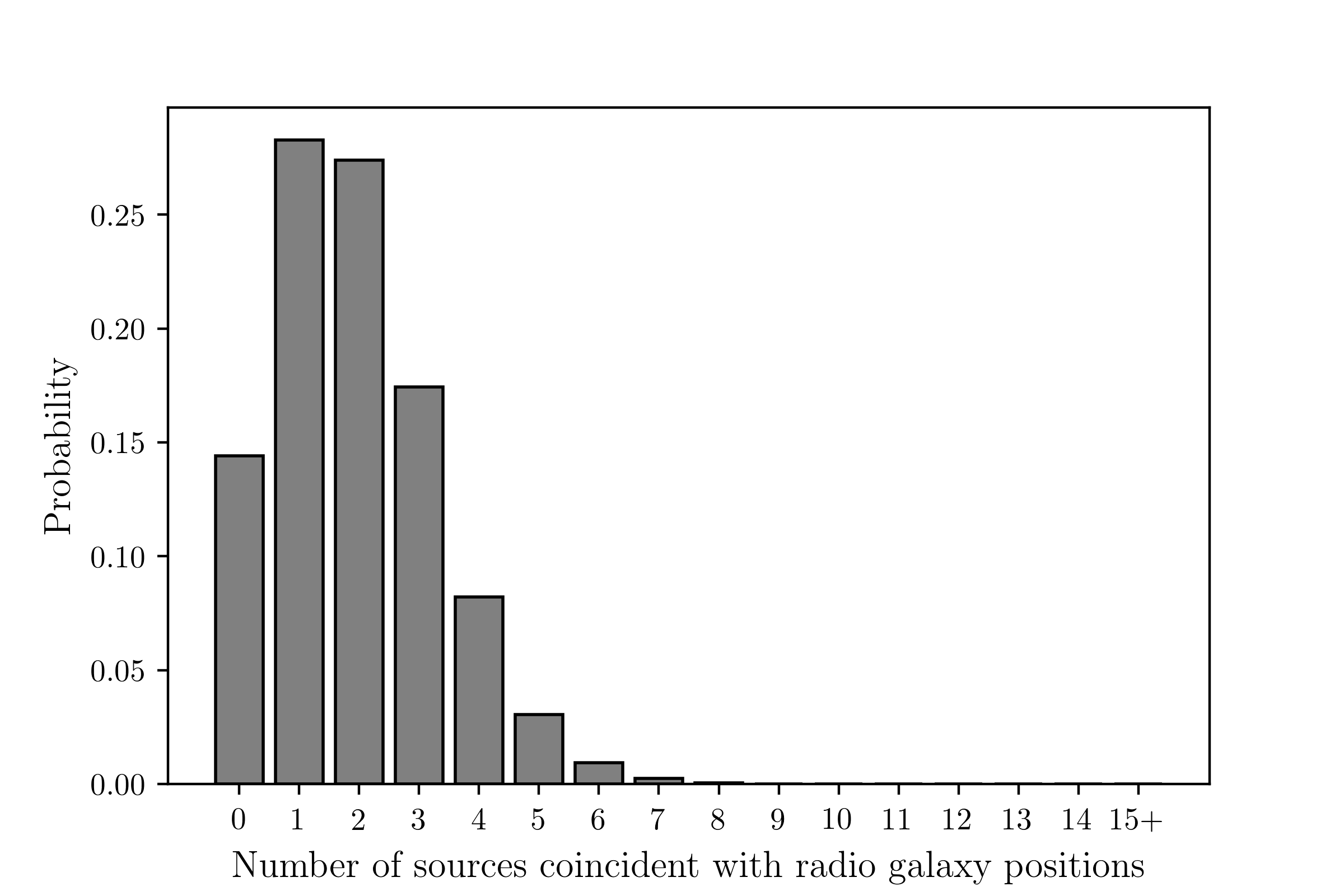}
    \caption{The probability distribution which characterizes the probability of $y$ number of chance correlations of spatially coincident sub-threshold $\gamma$-ray excesses with the positions of TRGs in our sample. The bins of $y = 15$ to $y = 72$ are combined, due to their very low probability.}
    \label{fig:lookelse}
\end{figure}

This probability distribution is shown in Figure \ref{fig:lookelse}, and we see that the probability of obtaining 4 STRGs through chance spatial correlation is 8\%, with the chance of at least 1 STRG being a false positive being 75\%.

\subsection{Sample Galaxy Core Dominance}
\label{sec:core}
The core dominance is a general measure of the intensity of the radio flux at the core of a galaxy with respect to the total radio emission of the galaxy. \cite{giovannini_vlbi_1994} suggests that measured core dominance of a galaxy could provide a measure of the orientation angle of jet, with a more dominant radio core indicating a smaller orientation angle. In blazars we are looking directly down the jet, and thus there is strong core dominance in this class of AGN. As $\gamma$-rays in blazars are intensified by Doppler boosting in the jets, if kpc scale jet emission in fact dominates the 4FGL galaxies in our sample (rather than large-scale processes) the detected galaxies will tend towards the higher end of the core dominance distribution.

We take the core dominance values from the VLA observations of the BCS quoted in \cite{liuzzo_bologna_2009} as these provide an almost complete representation of the radio galaxies in the BCS, although there are other ways of defining and calculating the core dominance, such as \cite{fan_core_2003}. The values from \cite{liuzzo_bologna_2009} correspond to the ratio between observed and estimated core radio power \citep{giovannini_vlbi_1994}, with a value greater than $1$ indicating a Doppler-boosted core, and less than $1$ indicating a deboosted core. VLA observations of the BCS are incomplete, but data are available for 57 galaxies in the sample (including all six 4FGL radio galaxies and two STRGs: 3C\,272.1 and B2\,1040+31). For a further 5 galaxies in our sample, there are upper limits to core dominance available; one of these is the STRG 3C\,293. Appendix Figure \ref{fig:core_dom} shows the distribution of core dominances for each radio galaxy.

Considering the 4FGL radio galaxies in the sample, 3C\,264 has a core dominance value of just above 1 indicating mild boosting. NGC\,2484, 4C\,39.12 and NGC\,315 all have core dominance values greater than 3, indicating stronger Doppler boosting in the core. This is indicative of strongly jet dominated $\gamma$-ray emission. The exceptional 4FGL radio galaxy is 3C\,274 (M\,87), where a core dominance $< 1$ indicates a deboosted core.

For the STRGs where we have values for core dominance, these are generally smaller than those of the 4FGL galaxies, with none of them above 1, although the sample size is small. Both 4C\,29.41 and 3C\,272.1 have core dominance values below 1, indicating that any emission from these radio galaxies may not originate from processes in the kpc scale jet, similar to the emission causing the spectral component seen in the core of Cen\,A, as discussed in \cite{brown_discovery_2017}, and searched for in \cite{rulten_search_2020} (for example, emission from a parsec scale inner jet, or an additional population of cosmic rays). In the case of 3C\,293, a 95\% confidence limit is in place of a measurement with a maximum value of 1.42. Thus it is difficult to conclude whether this radio galaxy is Doppler boosted or not.

The majority of our overall sample is boosted, with core dominance values greater than 1. This includes most of the $\gamma$-ray emitting 4FGL radio galaxies. The STRGs for which we have data are not generally among these, with only 3C\,293 possibly being mildly boosted.

\section{Discussion} 
\label{sec:discussion}

One motivating factor in choosing the BCS as the basis for our survey is that the redshifts and optical magnitudes of its radio galaxies are consistent with those of the 4FGL radio galaxies. Within this distribution the STRGs are neither particularly bright or close, so it is difficult to understand why we might see $\gamma$-ray emission from these 4 radio galaxies in particular. This reinforces the conclusion in Section \ref{sec:stats}, which suggests that the $\gamma$-ray emission is likely just a chance association.

The core dominance values described in Section \ref{sec:stats} indicate that the 4FGL radio galaxies have Doppler boosted jet dominant emission, which prevents us from observing any non-jet emission, should it be present. The only 4FGL radio galaxy in the BCS that is not core dominant is 3C\,274, better known as M\,87 (\citealt{abdo_fermi_2009}, \citealt{beilicke_discovery_2007}). M\,87 is a unique case, as it is a very massive galaxy at the centre of the Virgo supercluster and like Cen\,A it is close to Earth ($z = 0.0043$). As a deboosted radio galaxy, it is likely that some of the observed $\gamma$-ray emission occurs close to the black hole \cite{acciari_radio_2009} and not in the kpc jet, an hypothesis backed up by M\,87's variable nature. However, as the core and jet cannot be resolved from one another in $\gamma$-rays (unlike Cen\,A), it is more difficult to distinguish between jet and core emission. Furthermore, the spectral shape in M\,87 is variable and this makes it exceptionally difficult to identify persistent spectral features like the hardening observed in Cen\,A (which is non-variable on the timescale of the \textit{Fermi}-LAT mission,
with a variability index of 8.25 in the 4FGL). 

Comparing the core dominance values of the STRGs and the 4FGL radio galaxies, we find that (with the exception of 3C\,274/M\,87) these are not consistent with one another, as the 4FGL radio galaxies appear to have kpc-jet dominated emission, while the STRGs do not. Given that the STRGs are neither particularly bright nor particularly close, this further reinforces the conclusion that this apparent $\gamma$-ray emission is likely a chance association. 

\cite{angioni_gamma-ray_2019} find a correlation between radio core flux density and observed $\gamma$-ray luminosity in a sample of radio galaxies. However they also find other indicators suggesting that the observed $\gamma$-ray emission is not due to Doppler boosting in the kpc-jets, (such as a lack of correlation between $\gamma$-ray luminosity and core dominance) providing evidence for $\gamma$-ray emission which is not from the kpc scale jet. Figure \ref{fig:flux_cd} displays $\gamma$-ray flux plotted against core dominance; this is consistent with \cite{angioni_gamma-ray_2019} and \cite{rulten_search_2020} who find no correlation between $\gamma$-ray luminosity and core dominance in their analysis of the known 4FGL radio galaxies. Although these authors use a different method to calculate core dominance to \cite{liuzzo_bologna_2009}, which we use, all use radio measurements at similar frequencies (5 GHz and 8.4 GHz respectively). This supports the hypothesis that observable $\gamma$-ray emission is not necessarily linked to core dominance at these radio frequencies.

Assuming that the photon counts of the 4 STRGs grow linearly with time, we can estimate how much more \textit{Fermi}-LAT observations we would require for these objects to reach the $5 \sigma$ level. For the most significant of our 4 STRGs, B2\,0800+24, this is 6 years . Nonetheless, given the potentially variable nature of radio galaxies, it is worth monitoring the STRGs for changes in source significance and $\gamma$-ray flux.

\section{Conclusions} 
\label{sec:conclusions}
We executed an unbiased survey of radio galaxies with \textit{Fermi}-LAT targeting radio galaxies from the Bologna Complete Sample, with the hypothesis that any detectable $\gamma$-ray emission could be caused by mechanisms other than the jet. We find some $\gamma$-ray emission spatially coincident with 4 of these, but there is insufficient evidence to claim a detection. Furthermore, a calculation to understand the impact of the look-elsewhere effect on our survey suggests that this $\gamma$-ray emission is likely to be a chance correlation. This is supported by the fact that these 4 STRGs are generally unremarkable, giving no explanation as to why these 4 in particular might be observed out of the entire sample. We calculate flux limits for all of the TRGs and STRGs radio galaxies in our sample. 

The question of whether any $\gamma$-ray emission can be detected from non-jet processes in AGN therefore remains an open one. However, given the dominance of radio galaxies in the AGN population, further studies, using different catalogues and selection criteria, are warranted.

\section*{Acknowledgements}
The authors would like to acknowledge the excellent data and analysis tools provided by the NASA \textit{Fermi} collaboration, without which this work could not be done.  In addition, this research has made use of the NASA/IPAC Extragalactic Database (NED) which is operated by the Jet Propulsion Laboratory, California Institute of Technology, under contract with the National Aeronautics and Space Administration, the SIMBAD database, operated at CDS, Strasbourg, France and Montage, funded by the National Science Foundation.  We would also like to thank Matthew Capewell, Alastair Edge, Jamie Graham and Anthony Brown for useful discussions, and the referee for valuable feedback on this work. 

MH acknowledges funding from Science and Technology Facilities Council (STFC) PhD Studentship ST/S505365/1, and PMC and CBR acknowledge funding from STFC consolidated grant ST/P000541/1. 




\bibliographystyle{mnras}
\bibliography{BolognaSample.bib}



\appendix
\newpage
\onecolumn
\section{Radio Galaxy Sample}

\begin{table}[!ht]
\centering
\caption*{}
\begin{tabular}{ccccccc}
\hline \hline
Galaxy Name & Morphology & Redshift & RA ($\degree$) & Dec ($\degree$) & Flux Limit & $\gamma$-Rays Detected?\\
\hline
4C\,39.12 & FR\,0 & 0.0202 & 53.58 & 39.36 & $2.78 \times 10^{-6}$ & Yes, in 4FGL\\
4C\,31.04 & FR\,0 & 0.0592 & 19.9 & 32.18 & $6.00 \times 10^{-8}$ & No\\
B2\,0222+36 & FR\,0 & 0.0327 & 36.36 & 37.17 & $8.08 \times 10^{-8}$ & No\\
B2\,0648+27 & FR\,0 & 0.0409 & 53.58 & 27.46 & $8.35 \times 10^{-8}$ & No\\
4C\,30.19 & FR\,0 & 0.0909 & 160.13 & 29.97 & $2.83 \times 10^{-8}$ & No\\
B2\,1040+31 & FR\,0 & 0.036 & 160.83 & 31.52 & $8.77 \times 10^{-7}$ & STRG\\
N4278 & FR\,0 & 0.0021 & 185.03 & 29.28 & $4.76 \times 10^{-8}$  & No\\
B2\,1512+30 & FR\,0 & 0.0931 & 228.52 & 30.15 & $1.39 \times 10^{-8}$ & No\\
IC\,4587 & FR\,0 & 0.0442 & 239.97 & 25.94 & $2.76 \times 10^{-8}$ & No\\
B2\,1855+37 & FR\,0 & 0.0552 & 284.41 & 38.01 & $2.78 \times 10^{-8}$ & No\\
UGC\,367 & FR\,I & 0.0321 & 9.27 & 25.7 & $5.69 \times 10^{-8}$ & No\\
NGC\,326 & FR\,I & 0.0472 & 14.6 & 26.87 & $1.17 \times 10^{-8}$ & No\\
NGC\,315 & FR\,I & 0.0167 & 14.45 & 30.35 & $2.64 \times 10^{-6}$ & Yes, in 4FGL \\
3C\,31 & FR\,I & 0.0169 & 16.85 & 32.41 & $8.41 \times 10^{-8}$ & No\\
NGC\,507 & FR\,I & 0.0164 & 20.92 & 33.26 & $2.08 \times 10^{-8}$ & No\\
NGC\,708 & FR\,I & 0.016 & 28.19 & 36.15 & $4.53 \times 10^{-8}$ & No\\
4C\,35.03 & FR\,I & 0.0375 & 32.41 & 35.8 & $2.41 \times 10^{-8}$ & No\\
3C\,66B & FR\,I & 0.0215 & 35.8 & 42.99 & $2.15 \times 10^{-7}$ & No\\
3C\,76.1 & FR\,I & 0.0328 & 45.81 & 16.44 & $3.09 \times 10^{-8}$ & No\\
B2\,0326+39 & FR\,I & 0.0243 & 52.35 & 39.8 & $1.75 \times 10^{-8}$ & No\\
B2\,0708+32 & FR\,I & 0.0672 & 107.94 & 32.05 & $1.74 \times 10^{-8}$ & No\\
NGC\,2484 & FR\,I & 0.0413 & 119.6 & 37.78 & $1.29 \times 10^{-6}$ & Yes, in 4FGL\\
B2\,0800+24 & FR\,I & 0.0433 & 120.82 & 24.68 & $1.33\times 10^{-6}$ & STRG\\
4C\,32.26 & FR\,I & 0.068 & 130.3 & 32.42 & $2.07 \times 10^{-8}$ & No\\
B2\,0913+38 & FR\,I & 0.0711 & 139.17 & 38.68 & $9.79 \times 10^{-8}$ & No\\
B2\,0915+32 & FR\,I & 0.062 & 139.53 & 32.43 & $2.60 \times 10^{-8}$ & No\\
B2\,1102+30 & FR\,I & 0.072 & 166.35 & 30.16 & $4.27 \times 10^{-8}$ & No\\
4C\,29.41 & FR\,I & 0.0489 & 169.1 & 29.25 & $1.25 \times 10^{-6}$ & Yes, 4FGL\\
B2\,1116+28 & FR\,I & 0.0667 & 169.75 & 27.9 & $2.08 \times 10^{-8}$ & No\\
NGC\,3665 & FR\,I & 0.0067 & 171.18 & 38.76 & $5.91 \times 10^{-8}$ & No\\
3C\,264 & FR\,I & 0.0206 & 176.27 & 19.61 & $3.03 \times 10^{-6}$ & Yes, in 4FGL\\
B2\,1144+35 & FR\,I & 0.063 & 176.71 & 35.48 & $1.39 \times 10^{-8}$ & No\\
B2\,1204+24 & FR\,I & 0.0769 & 181.78 & 23.91 & $4.45 \times 10^{-8}$ & No\\
3C\,272.1 & FR\,I & 0.0037 & 186.27 & 12.89  & $1.04 \times 10^{-6}$ & STRG\\
3C\,274 (M 87) & FR\,I & 0.0043 & 187.71 & 12.39 & $1.31 \times 10^{-5}$ & Yes, in 4FGL \\
B2\,1243+26B & FR\,I & 0.0891 & 191.58 & 26.45 & $2.00 \times 10^{-8}$ & No\\
NGC\,4839 & FR\,I & 0.0246 & 194.35 & 27.5 & $5.58 \times 10^{-8}$ & No\\
NGC\,4869 & FR\,I & 0.0224 & 194.85 & 27.91 & $3.99 \times 10^{-8}$ & No\\
NGC\,4874 & FR\,I & 0.0239 & 194.9 & 27.96 & $4.31 \times 10^{-8}$ & No\\
4C\,29.47 & FR\,I & 0.0728 & 199.77 & 29.64 & $1.53 \times 10^{-8}$ & No\\
NGC\,5127 & FR\,I & 0.0161 & 200.94 & 31.57 & $5.53 \times 10^{-8}$ & No\\
NGC\,5141 & FR\,I & 0.0175 & 201.21 & 36.38 & $3.06 \times 10^{-8}$ & No\\
B2\,1339+26 & FR\,I & 0.0722 & 205.46 & 26.37 & $6.22 \times 10^{-8}$ & No\\
4C\,26.42 & FR\,I & 0.0633 & 207.29 & 26.59 & $3.77 \times 10^{-8}$ & No\\
3C\,293 & FR\,I & 0.0452 & 208.07 & 31.45 & $8.80 \times 10^{-7}$ & STRG\\
B2\,1357+28 & FR\,I & 0.0629 & 210.0 & 28.5 & $1.60 \times 10^{-8}$ & No\\
3C\,296 & FR\,I & 0.0237 & 214.22 & 10.81 & $8.41 \times 10^{-8}$ & No\\
4C\,25.46 & FR\,I & 0.0813 & 218.18 & 24.93 & $3.30 \times 10^{-8}$ & No\\
B2\,1441+26 & FR\,I & 0.0621 & 221.03 & 26.02 & $9.04 \times 10^{-9}$ & No\\

\hline

\end{tabular}
\end{table}

\newpage

\begin{table}[!ht]
\centering
\begin{tabular}{ccccccc}
\hline \hline
Galaxy Name & Morphology & Redshift & RA ($\degree$) & Dec ($\degree$) & Flux Limit & $\gamma$-Rays Detected?\\
\hline
3C\,305 & FR\,I & 0.041 & 222.34 & 63.27 & $4.06 \times 10^{-8}$ & No\\
3C\,310 & FR\,I & 0.054 & 226.24 & 26.02 & $6.28 \times 10^{-8}$ & No\\
B2\,1525+29 & FR\,I & 0.0653 & 231.94 & 28.92 & $1.92 \times 10^{-8}$ & No\\
B2\,1528+29 & FR\,I & 0.0843 & 232.54 & 29.01 & $3.20 \times 10^{-8}$ & No\\
NGC\,6086 & FR\,I & 0.0313 & 243.15 & 29.49 & $5.15 \times 10^{-8}$ & No\\
B2\,1613+27 & FR\,I & 0.0647 & 243.88 & 27.5 & $4.93 \times 10^{-8}$ & No\\
NGC\,6107 & FR\,I & 0.0296 & 244.33 & 34.9 & $3.49 \times 10^{-8}$ & No\\
NGC\,6137 & FR\,I & 0.031 & 245.76 & 37.92 & $1.79 \times 10^{-8}$ & No\\
3C\,338 & FR\,I & 0.0303 & 247.16 & 39.55 & $3.71 \times 10^{-8}$ & No\\
B2\,1637+29 & FR\,I & 0.0875 & 249.83 & 29.85 & $3.28 \times 10^{-8}$ & No\\
4C\,32.52 & FR\,I & 0.0631 & 254.75 & 32.49 & $6.02 \times 10^{-8}$ & No\\
4C\,30.31 & FR\,I & 0.0351 & 255.19 & 30.14 & $4.97 \times 10^{-8}$ & No\\
B2\,1736+32 & FR\,I & 0.0741 & 264.65 & 32.93 & $1.13 \times 10^{-8}$ & No\\
B2\,1752+32B & FR\,I & 0.0449 & 268.65 & 32.57 & $3.79 \times 10^{-8}$ & No\\
3C\,449 & FR\,I & 0.0181 & 337.84 & 39.36 & $3.47 \times 10^{-8}$ & No\\
B2\,2236+35 & FR\,I & 0.0277 & 339.62 & 35.33 & $3.72 \times 10^{-8}$ & No\\
3C\,465 & FR\,I & 0.0301 & 354.62 & 27.03 & $5.98 \times 10^{-8}$ & No\\
3C\,33 & FR\,II & 0.0595 & 17.21 & 13.31 & $1.83 \times 10^{-8}$ & No\\
3C\,98 & FR\,II & 0.0306 & 59.73 & 10.43 & $2.18 \times 10^{-8}$ & No\\
3C\,192 & FR\,II & 0.0597 & 121.36 & 24.16 & $3.04 \times 10^{-8}$ & No\\
4C\,32.15 & FR\,II & 0.0507 & 60.08 & 32.51 & $9.86 \times 10^{-8}$ & No\\
IC\,2402 & FR\,II & 0.0675 & 132.0 & 31.79 & $3.39 \times 10^{-8}$ & No\\
3C\,236 & FR\,II & 0.0989 & 151.51 & 34.9 & $1.69 \times 10^{-8}$ & No\\
3C\,277.3 & FR\,II & 0.0857 & 193.4 & 27.1 & $4.74 \times 10^{-8}$ & No\\
3C\,321 & FR\,II & 0.096 & 232.93 & 24.07 & $3.21 \times 10^{-8}$ & No\\
3C\,326 & FR\,II & 0.0895 & 238.04 & 20.09 & $4.07 \times 10^{-8}$ & No\\
3C\,382 & FR\,II & 0.0586 & 278.76 & 32.7 & $5.01 \times 10^{-8}$ & No\\
3C\,388 & FR\,II & 0.0917 & 281.01 & 45.56 & $1.42 \times 10^{-8}$ & No\\
3C\,390.3 & FR\,II & 0.0569 & 280.54 & 79.77 & $6.78 \times 10^{-7}$ & No\\
\hline
\end{tabular}
\caption{The radio galaxies selected from the BCS for use as targets in our search with the \textit{Fermi}-LAT data. The galaxy redshifts are from \protect\cite{giovannini_bologna_2005}, and the RA and Dec values are taken from the NASA/IPAC Extragalactic Database (NED), and quoted where appropriate to two decimal places. $E^{2} \frac{dN}{dE}$ is the energy multiplied by the differential flux for a bin of a spectral energy distribution, with units of $\mathrm{MeV} \mathrm{{cm}^{2-}} \mathrm{{s}^{-1}}$. To obtain the flux limits quoted, we calculate a single bin SED across our entire analysed energy range at the position of each TRG where no source is detected. Where a source is detected (in the case of the STRGs and 4FGL sources) we compute our upper limit for this source instead. }
\label{tbl:radgals}
\end{table}

\noindent  
\newpage
\section{Additional Figures}

\begin{figure}[!ht]
    \centering
    \includegraphics[width=500pt]{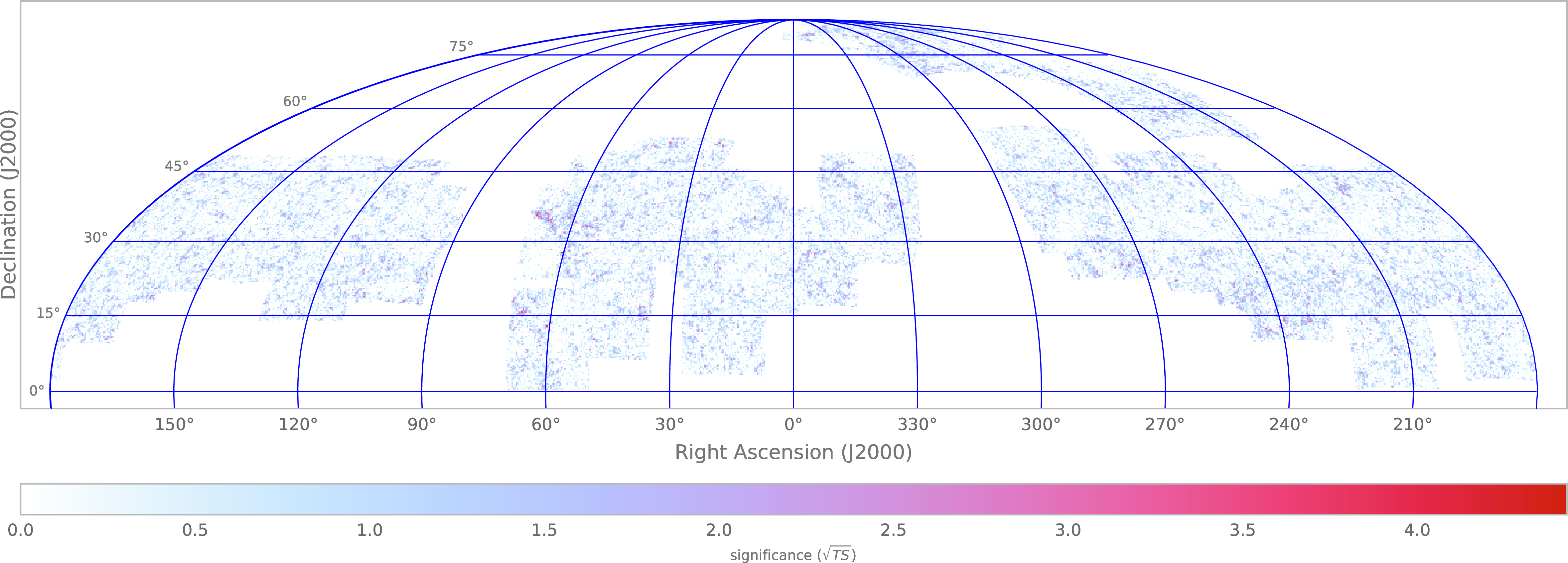}
    \caption{A TS map of all the analyzed ROIs across the northern hemisphere. Coloured regions indicate where observations and modelling have taken place, white regions have not been analysed. The colour scale shows the significance (square root of TS) of each bin, and largely shows random fluctuations from the Galactic and isotropic $\gamma$-ray diffuse backgrounds. As this map was generated \textit{after} the `find sources' algorithm, the upper limit on TS is approximately 16, as almost all of the points with an higher TS than this have been fitted with a point source. This map shows no significant areas of $\gamma$-ray excess that are not fitted by a model, so we can be confident that our modelling was successful.}
    \label{fig:ts_map}
\end{figure}

\begin{figure}[!ht]
    \centering
    \includegraphics[width=500pt]{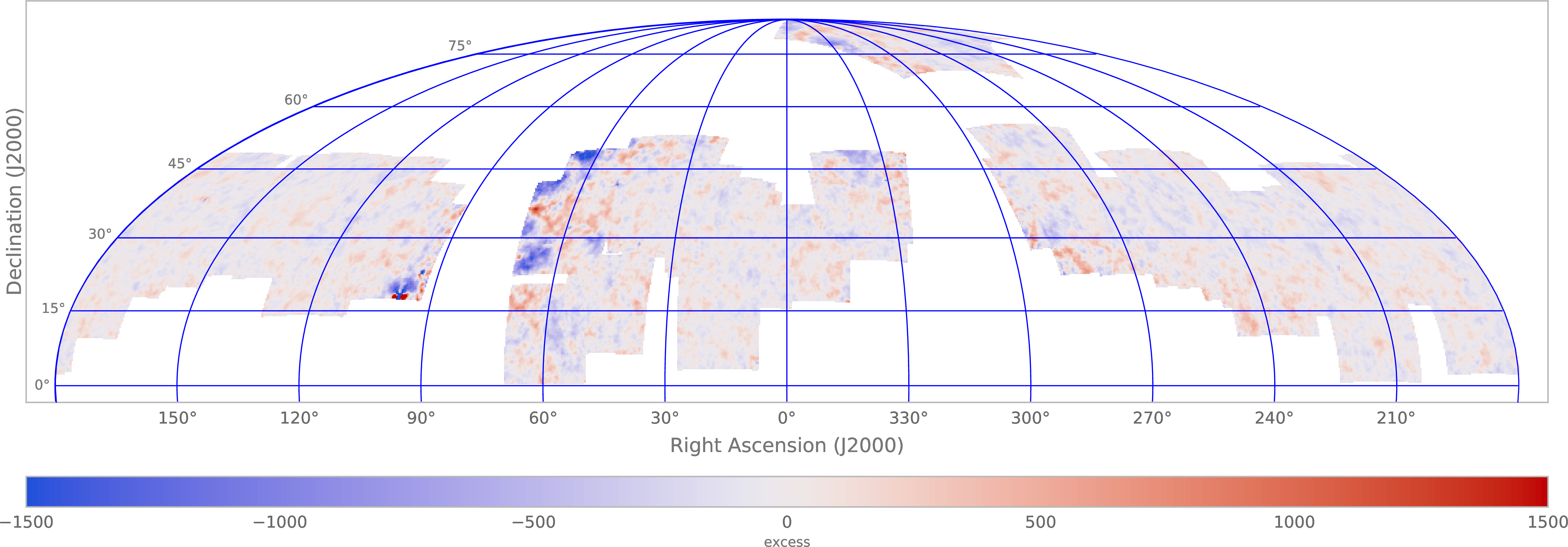}
    \caption{The residual map of all of the analyzed ROIs, across the northern hemisphere. Coloured regions indicate where observations and modelling have taken place, white regions show where there is no data. The colour scale shows residual counts, and consistently averages to around 0. The areas of poorer quality modelling adjacent to the left and right hand vertical white areas, are due to edge effects from the Galactic plane. This plot also shows that the ROIs overlap with one another, necessitating that we treat the TRGs independently. }
    \label{fig:res_map}
\end{figure}

\begin{figure}[!ht]
    \centering
    \includegraphics[width=500pt]{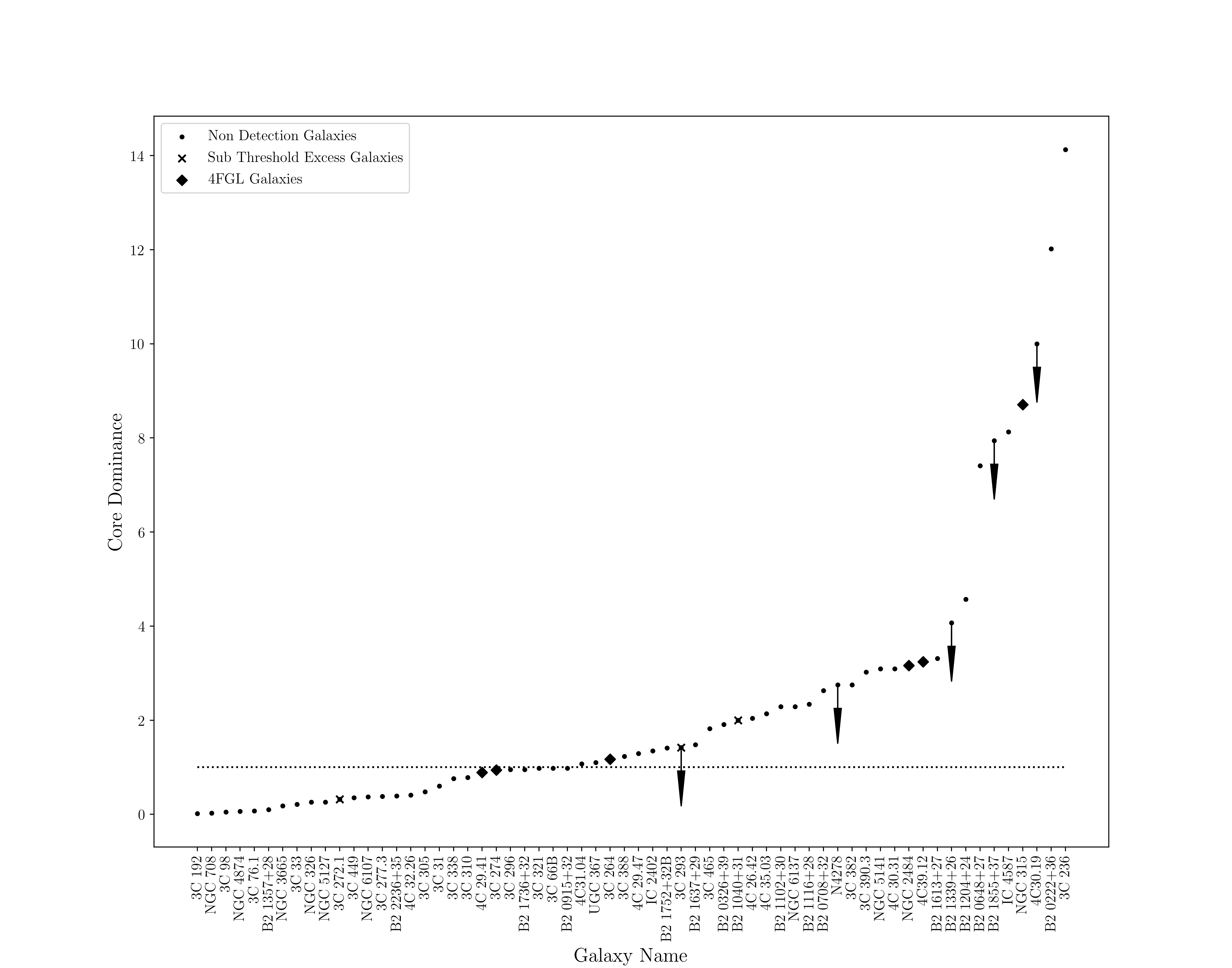}
    \caption{The distribution of the known core dominance values for the radio galaxies in the BCS, using values from \protect\cite{liuzzo_bologna_2009}. Not all of the radio galaxies in our sample have measurements of core dominance, and thus we cannot show a value for every TRG in the sample. Downward facing arrows indicate a 95\protect\% confidence limit, rather than an accurate measure of core dominance. The grey dashed line indicates a core dominance of 1. Radio galaxies above this line are therefore Doppler boosted in their jets, and those below are deboosted. Known 4FGL radio galaxies are indicated by a diamond and the STRGs with a cross.}
    \label{fig:core_dom}
\end{figure}

\begin{figure}[!ht]
    \centering
    \includegraphics[width=500pt]{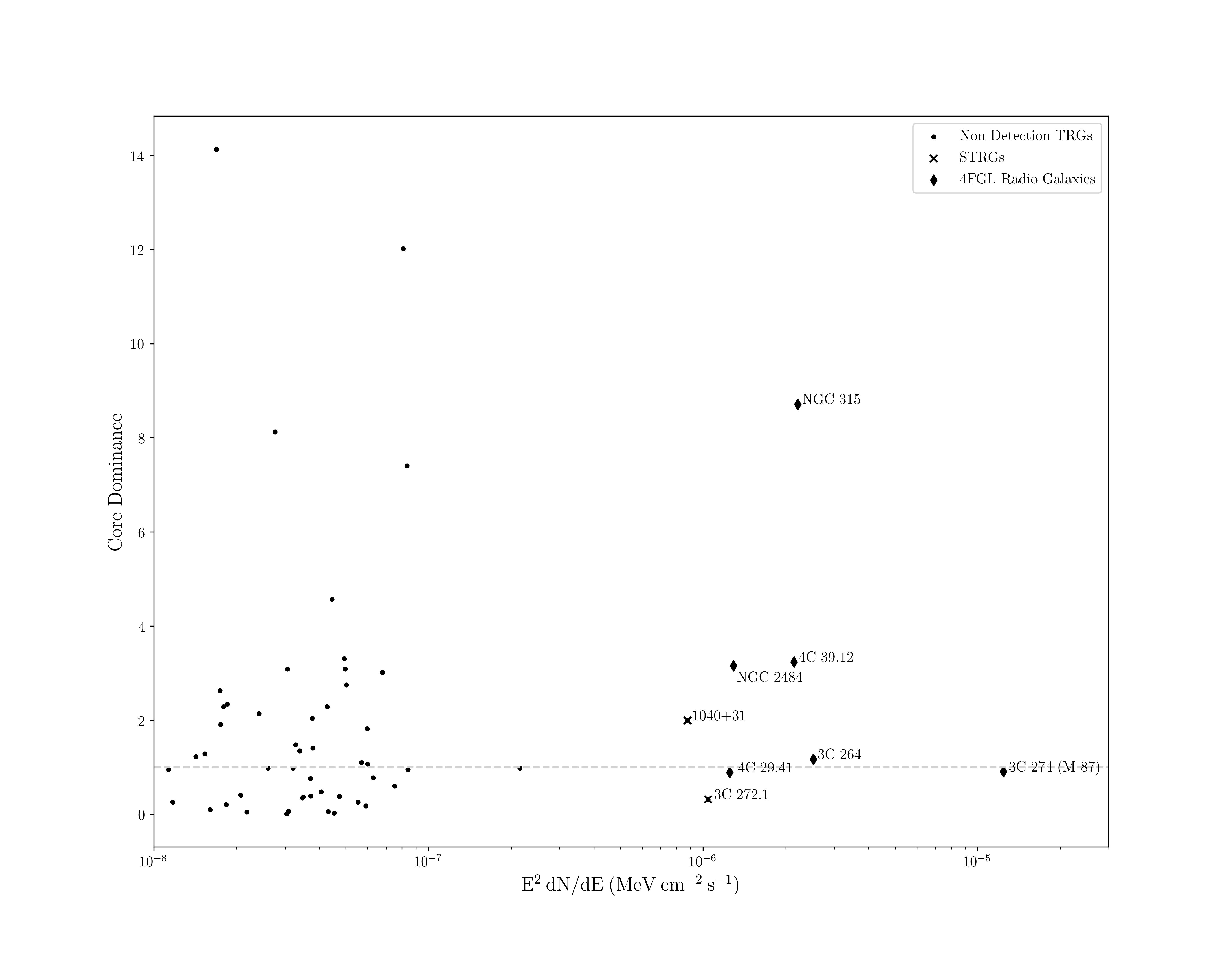}
    \caption{The distribution of known core dominance values for the radio galaxies in the BCS, plotted against the calculated $\gamma$-ray flux values shown in Table \ref{tbl:radgals}. All values are upper limits, except the significant detections 4C\,39.12, NGC\,315, 3C\,264 and 3C\,274 (M\,87) where the actual flux values are used. The grey dashed line indicates a core dominance of 1. Radio galaxies above this line are therefore Doppler boosted in their jets, and those below are deboosted. Known 4FGL radio galaxies are indicated by a diamond and the STRGs with a cross, and are labelled.}
    \label{fig:flux_cd}
\end{figure}

\bsp	
\label{lastpage}
\end{document}